\documentstyle[aps,12pt]{revtex}

\widetext
\draft
\tighten
\preprint{EFUAZ FT-98-56}

\begin{document}

\title{Comment on 'Comment on the Longitudinal Magnetic
Field of Circularly Polarized Electromagnetic Waves' by E. Comay
[Chemical Physics Letters 261 (1996) 601-604]\thanks{Submitted
to ``Chemical Physics Letters"}}

\author{{\bf Valeri V. Dvoeglazov}}

\address{
Escuela de F\'{\i}sica, Universidad Aut\'onoma de Zacatecas \\
Apartado Postal C-580, Zacatecas 98068, ZAC., M\'exico\\
Internet address:  valeri@cantera.reduaz.mx\\
URL: http://cantera.reduaz.mx/\~~valeri/valeri.htm}

\date{January 13, 1998}

\maketitle

\begin{abstract}
It is proved that necessary corrections in the Evans-Vigier modified
electrodynamics invalidate the arguments given by E. Comay [CPL {\bf 261}
(1996) 601]  against this model.  Moreover, from the conceptual viewpoint
Evans/Comay discussions in several journals contributed very little to the
modern electromagnetic theory due to many confusions of the both.
\end{abstract}


%\maketitle

\newpage

Recently, the model proposed by M. Evans and J.-P. Vigier~\cite{EV1}
was the object of the strong critics~\cite{COM1,COM2,COM3}. I cannot
consider the replies of M. Evans {\it et al.}~\cite{EVJF}
as sufficient ones. In fact, they contributed additional confusions
and misunderstandings to the discussion.\footnote{For instance, in
the reply by M. Evans and S. Jeffers in FPL~[5a]
the authors 1) considered relations which are valid if the circular
polarized radiation presents only; 2) in an attempt of a counterexample
they considered another path of integration and, in fact,
another type of radiation; and 3) contradicted the conclusions made
in ref.~\cite{Landau} without sufficient explanations.
So, in my opinion, their paper~[5a] is  irrelevant to the counterexamples
presented by E. Comay. This was pointed out by G. Hunter~\cite{Hunter}.}
This discussion
inspired me to express my own opinion on the problem of the longitudinal
modes of the electromagnetic field, see, e.~g.,
refs.~\cite{DVOE,DVOE1} and the present paper is the continuation of
my efforts to consider the problem {\it rigorously}. I have to mention
that I disagree with both E. Comay and M. Evans {\it et al.}

First of all, one should repeat
briefly what the authors of the cited works claimed. In ref.~\cite{EV1}
the longitudinal ``magnetic field" (an axial vector) ${\bf B}_\Pi \sim
{\bf E} \times {\bf E}^\ast$ was ascribed to a
circularly polarized electromagnetic wave. Moreover, in the subsequent
papers and books the {\bf B}- cyclic relations
\begin{equation}
{\bf B}^{(1)}\times {\bf B}^{(2)} = i B^{(0)} {\bf B}^{(3)\,\ast}
\quad \mbox{(et cyclic)} \label{1}
\end{equation}
were derived. The ${\bf B}^{(1)}$ and ${\bf B}^{(2)} = {\bf B}^{(1)\,\ast}$
are the accustomed transverse modes of the circularly polarized
electromagnetic wave (see  refs.~\cite{EV1,COM1} for detailed
explanation of the notation):
\begin{equation} {\bf B}^{(1)} =
{B^{(0)}\over \sqrt{2}} \pmatrix{i\cr 1\cr 0\cr} e^{i\phi}\quad,\quad {\bf
B}^{(2)} = {B^{(0)}\over \sqrt{2}} \pmatrix{-i\cr 1\cr 0\cr}
e^{-i\phi}\quad.\label{2}
\end{equation}
Thus, the longitudinal
phaseless component
\begin{equation}
{\bf B}^{(3)} = B^{(0)} \pmatrix{0\cr 0\cr 1\cr}
\end{equation}
of the ``magnetic field" in
the circular {\it complex} basis was defined there. This model
has come across the strong critics. E. Comay recently
argued that the model violates the relativistic covariance principle.
See~\cite{DVOE} for the discussion of whether this is
so.\footnote{As opposed to the opinion of E. Comay~[4, p. 252,
9th line from the bottom] the
principle of relativistic covariance means that the
physical laws expressed by equations preserve
their form in {\it any} frame.} Furthermore, on the basis of the
calculation of the line integral in the problem of rotating
dipole~\cite[p.228 of the Russian edition]{Landau}\footnote{In this
problem the polarization is defined by the direction ${\bf n}\cdot {\bf
d}$, ${\bf d}$ is the dipole moment~\cite[p.228 of the Russian
edition]{Landau}, thus giving the circular, elliptical and linear
polarizations when considering radiation emitted in various surface
angles. It was claimed by M. Evans (see the reference in~\cite{COM1}) that
the ${\bf B}^{(3)}$ is the property of the circular and, possibly,
elliptic polarizations and is equal to zero (??) in the linear
polarization (nevertheless, cf.~\cite{Lakh}). So, the problem noted by E.
Comay still may stands at the ${\bf B}^{(3)}$ theory, if one trusts the
Evans claims and if one considers the longitudinal field as a part of an
antisymmetric tensor of the second rank. Nevertheless, it is interesting
to note that, apart from the presence of different polarizations, the
energy flux is {\it not} isotropic in the particular example of Comay. It
depends on the polar angle $\theta$ as argued by Landau~\cite[p.228 of the
Russian edition]{Landau} even in the case of the consideration of
time-average flux over the period. All this may lead to further
speculations on the nature of ${\bf B}^{(3)}$.} the author of~\cite{COM1}
concluded that ``the flux of the electric field ${\bf E}$ through the area
{\it increases indefinitely} as time progresses.  It follows that if the
Maxwell equation in the vacuum $\bbox{\nabla} \times {\bf B} = \partial
{\bf E} /\partial t$ and (C)\footnote{(C) stands for the Evans' claim that
``the magnetic field  ${\bf B}^{(3)}$ is not associated with any real
electric field" which also may be doubted.} hold then the modified
electrodynamics leads to contradictions."

I agree. But, it is easy to show
that if one corrects the erroneous statement of M. Evans
that there cannot be any longitudinal components in the linear
polarized electromagnetic wave, then the path integral over the segment
$SR$ (see the figure 2 in~\cite{COM1}) does not vanish and it gives the
contribution to $\oint {\bf B}\cdot d{\bf l}$, which is equal in the
magnitude and opposite in the sign to that of the path segment $QP$. The
total path integral $\oint {\bf B}\cdot d{\bf l}$ is equal to zero, thus
invalidating the arguments by E. Comay.

For quantum field theorists it is known that the change of the
polarization state of massive particles can be made by the boost
(and/or other non-unitary operations).
On the other hand, it appears that for $j=1$ states (relevant to the
problem at hand) the change of polarization can be made by means of the
change of the basis of the corresponding {\it complex} vector space, {\it
i.~e.} by the rotation.  It is produced by an unitary matrix.  If one
describes the magnetic field as
\begin{equation}
{\bf B}^{\mbox{circ.}} =
{B^{(0)}\over \sqrt{2}} \left [\pmatrix{i\cr 1\cr 0\cr} e^{+i\phi} +
\pmatrix{-i\cr 1\cr 0\cr} e^{-i\phi} \right ]\, ,
\end{equation}
($\phi=\omega t -{\bf k}\cdot {\bf r}$)
on using
the unitary matrix
\begin{equation}
U ={1\over \sqrt{2}}\pmatrix{-i&1&0\cr
i&1&0\cr
0& 0&\sqrt{2}}
\end{equation}
one can obtain the linear polarized (in the
plane $XY$) radiation\footnote{If one wishes to see the real-valued
magnetic fields instead of phasors here they are:  \begin{equation} {\bf
B}_x^{\mbox{circ.}} = - \sqrt{2} B^{(0)} \sin\phi \quad , \, {\bf
B}_y^{\mbox{circ.}} = + \sqrt{2} B^{(0)} \cos\phi \quad  \, .
\end{equation}
or
\begin{equation}
{\bf B}_x^{\mbox{lin.}} =
+ B^{(0)} \cos\phi\quad,\, {\bf B}_y^{\mbox{lin.}} = + B^{(0)}
\cos\phi\quad,\,\label{rv}
\end{equation}
{\it i.~e.}, in the latter case one obtains the linear
polarized radiation with the polarization angle
equal to $\pi /4$ (defined by (\ref{rv})). Of
course, the given unitary matrix can be easily generalized
to account for other polarization angles.
Cf. with ref.~\cite[\S 7.2] {Jackson}.}$^{,}$\footnote{The
transformation of transverse components (\ref{2})
with the matrix $L$ used by G. Hunter is {\it not} generally unitary
(cf. with formulas (19) in~[7a]):
\begin{equation}
L\sim
\pmatrix{(A-B)\cos\alpha & -(A+B)\sin\alpha & 0\cr
(A-B)\sin\alpha & (A+B)\cos \alpha & 0\cr
0& 0& 1}\, ,
\end{equation}
with $\alpha$  being the polar angle of the cylindrical system of
coordinates. In the case of the linear polarization defined
in such a way~\cite{Hunter} one has ${\bf B}\times {\bf B}^\ast =
0$.  This transformation may also change the normalization of the
corresponding vectors which in the quantized case correspond to a particle
and an anti-particle. The determinant
of the transformation is, in general, {\it not} equal to the unit.  While
the determinant of our matrix is also not equal to the unit
($\mbox{det} U = -i$), but the norm of the corresponding
quantum states is still preserved (while this is not so for the
corresponding real quantities). By the way, Landau in \S 67 did not
work in terms of phasors; unfortunately, Dr. E. Comay did also not
elaborate this point. So, we do not know, what the definition of linear
polarized radiation does Landau imply in the problem of the rotating
dipole, presented by G. Hunter or presented by me in this work?
Nevertheless, cf. footnotes 3 and 5.}
\begin{equation} {\bf B}^{\mbox{lin.}} = U {\bf B}^{\mbox{circ.}} =
B^{(0)} \left [ \pmatrix{1\cr 0\cr 0\cr} e^{+i\phi}+\pmatrix{0\cr 1\cr
0\cr} e^{-i\phi} \right ]\ .  \end{equation} For this case of the linear
polarized radiation one has (instead of eq.  (\ref{1}))
\begin{equation}
B_x {\bf i} \times B_y {\bf j} = [B^{(0)}]^2 \pmatrix{0\cr 0\cr 1\cr} =
B^{(0)} B^{(0)} {\bf k}\, , \end{equation} {\it i.~e.}, the similar
relation to (\ref{1}), but already without the phase factor $e^{i\pi/2}$.
This conclusion is in the complete accordance with the Lakhtakia
consideration~[10a]: the Evans' `magnetostatic' field ${\bf B}_\Pi$ (or,
later, ${\bf B}^{(3)}$) ``may be defined for other than circularly
polarized plane waves".

These relations should be applied only in the {\it local} system,
which is connected with the observation point and the wave vector.
Otherwise, we come across big confusions. If one wishes
to use the {\it global} system of coordinates for this problem ${\bf B}
\sim \ddot {\bf d} \times {\bf n}$ is parallel to $OZ$ in the case of the
observation point in the plane $XY$; the vector cartesian components ${\bf
B}$ are already angular dependent, what makes the calculations to
be more difficult.

So, with necessary corrections the Evans-Vigier model
can be considered as useful and uncontradictory.\footnote{See,
nevertheless, the experimental controversy in
refs.~\cite{EXPER,RIKKEN,RAJA}.} In fact, the ${\bf B}$ cyclic relations
repeat tautologically the relations between spin components (after taking
into account the normalization), represent an interesting model, but
hardly to be considered as a fundamental theory (at the present level of
its development). Furthermore,
one should note that the ${\bf B}^{(3)}$ theory is
{\it not} the only candidate for the appear-to-be necessitated
generalization of the Maxwell's formalism.  As I am now aware the
longitudinal components of electromagnetic radiation were considered by
many authors in both XIX and XX centuries, e.~g., refs.~\cite{DVOE1,Chub}.
So, the common belief in the impossibility of existence of the
longitudinal electromagnetic-type interactions appears to me to be the
result of the greatest and uncomprehensible
mistake in the history of the XX
century science.  In my opinion, the most intriguing and promising theory
is the Weinberg $2(2j+1)$ component theory~\cite{Weinberg,DVA}, which also
represents the modified theory of electromagnetism~\cite{DVOE2} and~[20a].
The Weinberg theory was shown to be related to the problem of the
so-called Kalb-Ramond field~\cite{KR} (as well as the Evans-Vigier model).

In relation with all the above-mentioned I demonstrated in my recent works
(and this was let to know to Dr. Comay in 1995-1996) that:

\begin{itemize}

\item
The 3-vector ${\bf B}^{(3)}$ (which is defined
by (1)) may {\it not} be the entry of
the antisymmetric tensor field~\cite{DVOE};
it is {\it not} the $B_z \equiv F^{21}$
component but the entry of some
4-vector\footnote{This is obvious even from the
fact that ${\bf B}^{(3)}$
is a 3-vector (can possess three components, in the general case)
but $B_z$ is a number, the entry of $F^{\mu\nu}$, the electromagnetic
tensor. In the paper~\cite{DVOE} we used the
instant form of dynamics. It would be interesting to repeat
the calculations in the light-front form of relativistic
dynamics~\cite{Dirac}.}
provided that the Evans' definitions
for circularly polarized radiation are used.\footnote{It would be
still interesting to produce complete investigation of the
transformation properties of the cross products of transverse modes in
the case of various definitions of polarization states.} Lorentz
transformation rules for $(B^{(0)}, {\bf B}^{(3)})$ are the following:
\begin{mathletters}
\begin{eqnarray}
\label{lt1}
B^{(0)\,\prime} &=& \gamma (B^{(0)} -{\bbox\beta}\cdot {\bf B}^{(3)}
)\quad,\\
\label{lt2}
{\bf B}^{(3)\,\prime} &=& {\bf B}^{(3)} + {\gamma -1 \over
\beta^2} ({\bbox \beta} \cdot {\bf B}^{(3)}) {\bbox\beta} - \gamma
{\bbox\beta} B^{(0)} \quad, \
\end{eqnarray} 
\end{mathletters}
with ${\bbox\beta} ={\bf v}/c$\,,\,\,$\beta = \vert {\bbox\beta}\vert
= \mbox{tanh} \phi$\,,\, $\gamma ={1\over \sqrt{1-{\bbox\beta}^2}}
=\mbox{cosh} \phi$, and $\phi$ is the parameter of the Lorentz boost.

\item
Due to the previous item {\it there are no any reasons} that the quantity
which is {\it not}  a part of the antisymmetric tensor field $F^{\mu\nu}$
satisfies the Maxwell's vacuum equations $\partial_\mu F^{\mu\nu} =
0\,\, , \mu = 0,1,2,3$.

\item
In~\cite{DVOE2} the Weinberg-Tucker-Hammer
equation~[16a] and~\cite{Tuck}
\begin{equation}
\left [\gamma_{\alpha\beta} p_\alpha p_\beta +p_\alpha p_\alpha +2m^2
\right ] \Psi (x^\mu) =0\,
\end{equation}
($p_\alpha = -i\partial_\alpha$ and the euclidean metric being used)
was considered
on using the interpretation of the Weinberg $j=1$
field functions as $\Psi (x^\mu) = \mbox{column} (\chi \quad \varphi )$,
$\chi={\bf E}+i{\bf B}$, \, $\varphi = {\bf E}-i{\bf B}$. As a result we
arrive at the set of equations
\begin{eqnarray} \lefteqn{\left [ E^2 -
{\bf p}^{\,2}\right ]_{ij}({\bf E}^j+i{\bf B}^j)^{\parallel} -m^2 ({\bf
E}^i-i{\bf B}^i)^{\parallel} +\nonumber}\\
&+&\left [ E^2 +{\bf p}^{\,2}-
2 E ({\bf J}\cdot{\bf p})\right ]_{ij} ({\bf E}^j+i{\bf B}^j)^{\perp} -
m^2 ({\bf E}^i-i{\bf B}^i)^{\perp} =0\quad,\label{cl1}
\end{eqnarray}
and
\begin{eqnarray}
\lefteqn{\left [ E^2 -{\bf p}^{\,2}\right ]_{ij}({\bf E}^j-i{\bf
B}^j)^{\parallel} -m^2 ({\bf E}^i+i{\bf B}^i)^{\parallel} +\nonumber}\\
&+&\left [ E^2 + {\bf p}^{\,2}+2 E ({\bf J}\cdot{\bf p})\right ]_{ij}
({\bf E}^j-i{\bf B}^j)^{\perp} - m^2 ({\bf E}^i+i{\bf B}^i)^{\perp} =0
\quad.\label{cl2} \end{eqnarray}
One can see that  in the classical field
theory antisymmetric tensor fields are the fields with both transverse
and longitudinal components in the massless limit. The longitudinal parts
of the above equations do not contain the terms as $({\bf J}\cdot {\bf
p})$ provided that the longitudinal modes are associated with the plane
waves too. This can be easily seen on choosing the spin basis where
$(J^i)_{jk} = -i\epsilon^{ijk}$ and on using the definition of the
longitudinal modes, ${\bf p} \times ({\bf E}\pm i{\bf B})^\parallel \equiv
0$.  So, the Weinberg-Tucker-Hammer equations for antisymmetric tensor
fields (which are deduced on the basis of the general principles for
deriving relativistic equations) may describe the longitudinal components
with non-zero energy.

\item
If one considers the Maxwell's equations as the definitions for currents
and charges one arrives at the additional equations~[20a]:
\begin{mathletters}
\begin{eqnarray}
{\partial {\bf J}_e \over \partial t}+\mbox{grad} \rho_e &=&m^2 {\bf E}
\quad,\quad \mbox{curl} {\bf J}_m =0\, ,\\
{\partial {\bf J}_m \over \partial t} +\mbox{grad} \rho_m &=& 0\quad,\quad
\mbox{curl} {\bf J}_e = -m^2 {\bf B}\, ,
\end{eqnarray}
\end{mathletters}
$c=\hbar =1$ and the indices $e,m$ denotes electric and magnetic parts
respectively. They might be relevant to the old Einstein idea of the
dequantization of the charge and invoke immediately the additional concept
of the scalar chi-functions of boundary and initial conditions.  The
massless limit is easily found from these formulas.

\item
In the recent paper~[20d] we considered the general case of $2(2j+1)$
component field functions and 4-vector potential in the instant form of
relativistic dynamics (cf. with~\cite{DVA00}).
The cross products of   magnetic fields of different spin states
in the momentum representation (such as
${\bf B}^{(+)} ({\bf p}, \sigma) \times {\bf B}^{(-)} ({\bf p},
\sigma^\prime)$) may {\it not} be equal to zero and may be expressed  by
the ``time-like" potential and/or the gauge part of 3-potentials
for different spin states (also in the momentum representation):
\begin{mathletters}
\begin{eqnarray}
{\bf B}^{(+)} ({\bf p}, +1) \times {\bf B}^{(-)} ({\bf p}, +1)
&=& - {iN^2 \over 4m^2} p_3 \pmatrix{p_1\cr p_2\cr
p_3\cr} = -{\bf B}^{(+)} ({\bf p}, -1) \times {\bf B}^{(-)} ({\bf p},
-1)\,\, ,\\
{\bf B}^{(+)} ({\bf p}, +1) \times {\bf B}^{(-)} ({\bf p}, 0)
&=& - {iN^2 \over 4m^2} {p_r \over \sqrt{2}} \pmatrix{p_1\cr
p_2\cr p_3\cr} = + {\bf B}^{(+)} ({\bf p}, 0) \times {\bf B}^{(-)} ({\bf
p}, -1)\,\, ,\\
{\bf B}^{(+)} ({\bf p}, -1) \times {\bf B}^{(-)} ({\bf p}, 0)
&=& - {iN^2 \over 4m^2} {p_l \over \sqrt{2}} \pmatrix{p_1\cr
p_2\cr p_3\cr} = + {\bf B}^{(+)} ({\bf p}, 0) \times {\bf B}^{(-)} ({\bf
p}, +1)\,\, .
\end{eqnarray}
\end{mathletters}
$N$ is the normalization term; $p_{r,l} =p_1 \pm i p_2$.
Other cross products are equal to zero. Cf. with the formulas
(15a,15b,22)  in~[20d].

\end{itemize}

Concluding, on the basis of this my paper and previous ones I can state
that the possible existence of longitudinal components of antisymmetric
tensor field (and/or 4-potentials) does not contradict the principle of
relativistic covariance (but can still be related to the
action-at-a-distance concept and topological theories); the $\mbox{curl}$
of longitudinal components may satisfy the Maxwell equation after the
necessary modifications of the claims made by M. Evans {\it et al.}, but
this is not too necessary, because one can consider ${\bf B}^{(3)}$ to be
the longitudinal components of 4-vector potentials (and/or of the
polarization vector~\cite{Hunter}), which need not already to satisfy the
Maxwell equations for strengths. Finally, we found {\it two} ways for the
definition of the linear polarized radiation; this freedom is related to
the unobservability of phasors --- only real electric/magnetic fields are
observable in the present-day classical electrodynamics.

\acknowledgments
I am thankful to Profs. A. Chubykalo, G. Hunter, R. Kiehn and Y. S. Kim
for valuable discussions. I acknowledge many internet
communications of Dr. M. Evans  (1995-96) on the concept of the ${\bf
B}^{(3)}$ field, while frequently do {\it not} agree with him. I am
delighted by the dedication of Dr. E. Comay, which permitted me to express
my opinion on these matters once again.  I consider the Comay's critics
while irrelevant and unreasonable but provocative: in fact, it is aimed at
the indication of some tune points in the Evans-Vigier model, which are
required clarifications and corrections. Therefore, I  want to thank Dr.
E.  Comay too.

I am grateful to Zacatecas University for a professorship.
This work has been supported in part by the Mexican Sistema
Nacional de Investigadores, the Programa de Apoyo a la Carrera
Docente and by the CONACyT, M\'exico under the research
project 0270P-E.

%\begin{references}

\newpage

\begin{center}
{{\large \bf Reply to M. W. Evans}}\\

\bigskip

{\bf Valeri V. Dvoeglazov}\\

\medskip

{\it Escuela de F\'{\i}sica, UAZ\\
Apartado Postal C-580\\
Zacatecas 98068 Zac. M\'exico}

\end{center}

\bigskip
%\bigskip

Evans raised objections to my previous paper [5]
on the basis that it does not use the non-Abelian Stokes
theorem for proving existence of the ${\bf B}^{(3)}$ field.
Non-Abelian Stokes theorem indeed has been proposed
in the works of several authors, e.g. [1]; it is well known
and it represents equations of {\it  isospin} components.
The non-Abelian generalizations of other laws
of electrodynamics also present, e.g. [2].

However, I did not find, unfortunately, any correct form
of the non-Abelian Stokes theorem neither in Evans'
work [3],  which he refers to, nor in any other works of that
group.  To the best of my knowledge,
a correct connection between  spin and isospin has  not yet
been established.  Therefore, I consider that Evans' critical comment
[4]  of my papers [5] has no content and it is politically motivated.
It clarifies almost nothing in his own discussion with
Comay and Hunter. Comay's answer on my work [5a] is even
more irrelevant to the essence of the problem, see [5b].

%\bigskip

%{\bf REFERENCES.}\\

\newpage

\begin{center}

{{\large {\bf Reply to G. W. Bruhn}}}

\bigskip

{\bf V. V. Dvoeglazov}\\

\medskip

{\it Universidad de Zacatecas, Ap. Postal 636, Suc. 3 Cruces\\
Zacatecas 98064 Zac., M\'exico\\
E-mail: valeri@planck.reduaz.mx}

\end{center}

\medskip

\small{{\bf Abstract.} I show that the Bruhn critics of my article is based on his misunderstandings.
The major part of the critics of the Evans articles has been first given in my works
and in private communications to them.}

\bigskip

In ref.~\cite{Bruhn} G. W. Bruhn claims that he found errors in my articles~\cite{DV1,DV2}.
For instance, the following statements have been given by Bruhn: 
``Since ${\bf B}^{(3)}$
is merely the longitudinal component of a field ${\bf B}$ that has an additional transversal component ${\bf B}_\perp$ Dvoeglazov's result  contradicts the well-known Lorentz transform of the electromagnetic field
 where the longitudinal component remains unchanged." ``This result proves that V. V. Dvoeglazov Equ. (11b) cannot be true".

However, in fact, I called frequently the ${\bf B}^{(3)}$ to be ``the so-called magnetic field", just see the inverted commas in
the words "magnetic field" (the first line, the  page 228 of~\cite{DV2}).
Next (see the 5th line of the page 230 of~\cite{DV2}, just before the equation (11b), which Bruhn doubts)
I state explicitly ``The 3-vector ${\bf B}^{(3)}$ (which is defined by (1) [by Evans indeed]) may {\it not} be the entry of the antisymmetric tensor field; it is \ldots the entry of some 4-vector provided that  the Evans' definitions for circularly polarized radiation are used." I hope that Bruhn knows that 
the Lorentz transformation laws are different for an antisymmetric tensor field and a 4-vector field. The equation (11b) of the cited paper is precisely the transformation law for the 3-part of a 4-vector. Its parity properties have been dicussed in the discussion with Comay and Evans, cf. with ref.~[4a]. 

Moreover, in ref.~\cite{DV1} (published previously than~\cite{DV2}), see Eqs. (9), I proved the statement by explicite mathematical calculations. I again explicitely claimed: ``\ldots$B^{(0)}$ transforms as zero-component of the 4-vector and ${\bf B}^{(3)}$ as the space components of the 4-vector\ldots".

Finally, a quite inaccurate statement is given in the end of the Bruhn paper. The $SO(3)$ group is the
subgroup of the Lorentz group. So, it is obvious that $SO(3)$ symmetry  (but the different one from that 
given  by M. W. Evans) is compatible with the Lorentz covariance.

Thus, in my opinion, the Bruhn paper is some sort of diffamation.\footnote[1]{Unfortunately, the same sins can be found in the papers and the correspondence by M. W. Evans and E. Comay. So, I am {\bf not} going to enter 
in discussions with them in the future.}
While I acknowledge the trivial errors in the papers by M. W. Evans, E. Comay and G. W. Bruhn, I continue
to state that there are NO any calculational errors in my papers~\cite{DV1,DV2,DV3}.


\begin{thebibliography}{99}


\bibitem{EV1} M. W. Evans, {\it Physica} B{\bf 182} (1992) 227; ibid.
237; M. W. Evans and J.-P. Vigier, {\it The Enigmatic Photon.}
Vols.  1-3 (Kluwer Academic, Dordrecht, 1994-96), the third
volume with S.  Jeffers and S. Roy.

\bibitem{COM1} E. Comay, {\it Chem. Phys. Lett.} {\bf 261} (1996) 601.

\bibitem{COM2} E. Comay, {\it Physica} B{\bf 222} (1996) 150.

\bibitem{COM3} E. Comay, {\it Found. Phys. Lett.} {\bf 10} (1997) 245.

\bibitem{EVJF} M. W. Evans and S. Jeffers, {\it Found. Phys. Lett.}
{\bf 9} (1996) 587; M. W. Evans, {\it Found. Phys. Lett.} {\bf 10}
(1997) 255; ibid. (1997) 403.

\bibitem{Landau} L. D. Landau and E. M. Lifshitz, {\it Teoriya polya.}
(Nauka, Moscow, 1973) [English translation: {\it The Classical Theory
of Fields.} (Pergamon, Oxford, 1975)].

\bibitem{Hunter} G. Hunter, {\it The ${\bf B}^{(3)}$ Field: an
Assessment.} Preprint, July 1997; {\it The ${\bf B}^{(3)}$ Field
Controversy.} Preprint, July 1997.

\bibitem{DVOE} V. V. Dvoeglazov,  {\it Found. Phys. Lett.} {\bf 10} (1997)
383 (This paper presents itself a comment on the debates between E. Comay
and M. Evans, ref.~[4,5b] and it criticizes both authors).

\bibitem{DVOE1} V. V. Dvoeglazov,  in {\it The Enigmatic Photon.}
Vol.  IV, eds. M. W. Evans, G. Hunter, S. Roy and J.-P. Vigier
(Kluwer Academic, Dordrecht, 1997), Chapter 12.

\bibitem{Lakh} A. Lakhtakia, {\it Physica} B{\bf 191} (1993) 362;
D. M. Grimes, ibid {\bf 191} (1993) 367.

\bibitem{Jackson} J. D. Jackson, {\it Electrodin\'amica Cl\'asica.}
Spanish edition (Alhambra S. A., 1980).

\bibitem{EXPER} J. P. van der Ziel, P. S. Pershan and L. D. Malmstrom,
{\it Phys Rev. Lett.} {\bf 15} (1965) 190; {\it Phys. Rev.} {\bf 143}
(1966) 574.

\bibitem{RIKKEN} G. L. J. A. Rikken, {\it Optics Lett.}
{\bf 20}  (1995) 846; M. W. Evans, {\it Found. Phys. Lett.} {\bf 9}
(1996) 61.

\bibitem{RAJA} M. Y. A. Raja {\it et al.}, {\it Appl. Phys. Lett.} {\bf
67} (1995) 2123; {\it Appl. Phys.} B{\bf 64} (1997) 79.

\bibitem{Chub} A. E. Chubykalo and R. Smirnov-Rueda, {\it Mod. Phys.
Lett.} A {\bf 12} (1997) 1.

\bibitem{Weinberg} S. Weinberg, {\it Phys. Rev.} B{\bf 133} (1964) 1318;
ibid B{\bf 134} (1964) 882; ibid {\bf 181} (1969) 1893.

\bibitem{DVA} D. V. Ahluwalia and D. J. Ernst, {\it Mod. Phys. Lett.}
A{\bf 7} (1992) 1967;  D. V. Ahluwalia, M. B. Johnson and T. Goldman,
{\it Phys.  Lett.}  B{\bf 316} (1993) 102.

\bibitem{DVOE2} V. V. Dvoeglazov, {\it Helv. Phys. Acta} {\bf 70} (1997)
677, ibid. 686, ibid. 697.

\bibitem{KR} V. I. Ogievetskii and I. V. Polubarinov, {\it Sov. J. Nucl.
Phys.} {\bf 4} (1967) 156; K. Hayashi, {\it Phys. Lett.} B{\bf 44} (1973)
497; M.  Kalb and P.  Ramond, {\it Phys. Rev.} D{\bf 9} (1974) 2273.

\bibitem{DVOE3} V. V. Dvoeglazov, Preprints EFUAZ-FT
hep-th/9410174, hep-th/9604148, hep-th/9611068, hep-th/9712036,
hep-ph/9801287 (1994-98), submitted.

\bibitem{Dirac} P. A. M. Dirac, {\it Rev. Mod. Phys.} {\bf 21} (1949) 392.

\bibitem{Tuck} R. H. Tucker and C. L. Hammer, {\it Phys. Rev. D}{\bf 3}
(1971) 2448.

\bibitem{DVA00} D. V. Ahluwalia and D. J. Ernst, {\it Int. J. Mod. Phys.
E}{\bf 2} (1993) 397.

%\end{references}

\end{thebibliography}

\begin{thebibliography}{99}

[1] I. Aref'eva, Theor. Math. Phys. {\bf 43}, 353 (1980); N. E. Brali\'c,
Phys. Rev. D{\bf 22}, 3090 (1980); M. B. Mensky, Lett. Math. Phys.
{\bf 3}, 513 (1979); P. M. Fishbane, S. Gaziorowicz and P. Kaus, Phys. Rev.
D{\bf 24}, 2324 (1981); L. Di\'osi, Phys. Rev. D{\bf 27}, 2552 (1983);
J. Szczesny, Acta Phys. Polon. B{\bf 18}, 707 (1987);
Yu. A. Simonov, Sov. J. Nucl. Phys. {\bf 50}, 134 (1989);
B. Broda, J. Math. Phys. {\bf 33}, p. 1511 (1992);
in {\it Advanced Electromagnetism: Foundations,
Theory Applications.} Eds. T. W. Barrett and D. M. Grimes (Singapore,
World Sci., 1995),  p. 496-505;
D. Diakonov and V. Petrov, hep-th/9606104; F. A. Lunev,
Nucl. Phys. B{\bf 494}, 433 (1997); M. Hirayama and S. Matsubara,
Prog. Theor. Phys. {\bf 99}, 691 (1998); V.I. Shevchenko and Yu.A. Simonov,
hep-th/9802134; M. Hirayama, M. Kanno, M. Ueno and H. Yamakoshi,
Prog.Theor.Phys. {\bf 100}, 817 (1998); V.I. Shevchenko and Yu.A. Simonov,
Phys. Lett. B{\bf 437}, 146 (1998); R. L. Karp, F. Mansouri and J. S. Rno,
hep-th/9903221; M. Faber, A.N. Ivanov, N.I. Troitskaya and M. Zach,
hep-th/9907048; M. Hirayama and M. Ueno, hep-th/9907063;
R. L. Karp, F. Mansouri and J.S. Rno, J. Math. Phys. {\bf 40}, 6033 (1999);
K-I. Kondo and Y. Taira, hep-th/9911242; R. L. Karp and F. Mansouri,
hep-th/0002085, and references therein.\\

[2] A. Mukherjee, Z. Phys. C{\bf 32}, 619 (1986); K. Harada and
I. Tsutsui, Z. Phys. C {\bf 41}, 65 (1988);
J-G. Zhou, S-M. Li and Y-Y Liu, Phys. Rev. D{\bf 48}, 961 (1993);
P. Majumdar and H.S. Sharatchandra, Phys. Rev. D{\bf 58}, 067702 (1998);
P. Majumdar and H.S. Sharatchandra, Nucl.Phys.Proc.Suppl., {\bf 73},
620-622 (1999), and references therein.\\

[3] M. W. Evans et al., Apeiron, {\bf 6}, 222-226 (1999).\\

[4] M. W. Evans, Apeiron, {\bf 7}, 116-119 (2000).\\

[5] V. V. Dvoeglazov, Apeiron, {\bf 6}, 227-232 (1999), physics/9801024;
see also  physics/9907048.\\

\end{thebibliography}

\begin{thebibliography}{99}

\bibitem{Bruhn} G.~W.~Bruhn,~http://www.mathematik.tu-darmstadt.de/\~{}bruhn/EM-Lorentz-Transform.html .
In the similar articles published in the journals Bruhn somehow removed the references to my works.
However, it is still on his website.  I was not aware about the  Bruhn critics existence for a long time.

\bibitem{DV1} V. V. Dvoeglazov, Found. Phys. Lett., {\bf 10}, No. 4, pp.383-391 (1997),  physics/9611009 .

\bibitem{DV2} V. V. Dvoeglazov, Apeiron, {\bf 6}, No. 3-4, pp. 227-232 (1999), physics/9801024 .  

\bibitem{DV3} V. V. Dvoeglazov, Found. Phys. Lett., {\bf 13}, No. 4, pp. 387-393 (2000), physics/9907048; 
V. V. Dvoeglazov and J. L. Quintanar Gonzalez, Found. Phys. Lett.,   {\bf 19}, No. 2, 
195-200  (2006),  physics/0410169 .


\end{thebibliography}
\end{document}